\documentclass[a4paper]{article}

\usepackage{INTERSPEECH2021}
\usepackage{hyperref,amssymb,amsmath,graphicx,ctable,makecell,multirow,subcaption}
\usepackage{cite}
\usepackage{enumerate}

\title{Raw Waveform Encoder with Multi-Scale Globally Attentive Locally Recurrent Networks for End-to-End Speech Recognition}
\name{Max W. Y. Lam$^1$, Jun Wang$^1$, Chao Weng$^1$, Dan Su$^1$, Dong Yu$^2$}
\address{
  $^1$Tencent AI Lab, Shenzhen, China\\
  $^2$Tencent AI Lab, Bellevue WA, USA}
\email{\{maxwylam, joinerwang, cweng, dansu, dyu\}@tencent.com}

\begin{document}

\maketitle
\begin{abstract}
End-to-end speech recognition generally uses hand-engineered acoustic features as input and excludes the feature extraction module from its joint optimization. To extract learnable and adaptive features and mitigate information loss, we propose a new encoder that adopts globally attentive locally recurrent (GALR) networks and directly takes raw waveform as input. We observe improved ASR performance and robustness by applying GALR on different window lengths to aggregate fine-grain temporal information into multi-scale acoustic features. Experiments are conducted on a benchmark dataset \textit{AISHELL-2} and two large-scale Mandarin speech corpus of $5,000$ hours and $21,000$ hours. With faster speed and comparable model size, our proposed multi-scale GALR waveform encoder achieved consistent character error rate reductions (CERRs) from $7.9\%$ to $28.1\%$ relative over strong baselines, including Conformer and TDNN-Conformer. In particular, our approach demonstrated notable robustness than the traditional handcrafted features and outperformed the baseline MFCC-based TDNN-Conformer model by a $15.2\%$ CERR on a music-mixed real-world speech test set.
\end{abstract}

\begin{figure*}[ht!]
 \centering
\vspace{-0.1cm}
    \includegraphics[width=0.836\linewidth]{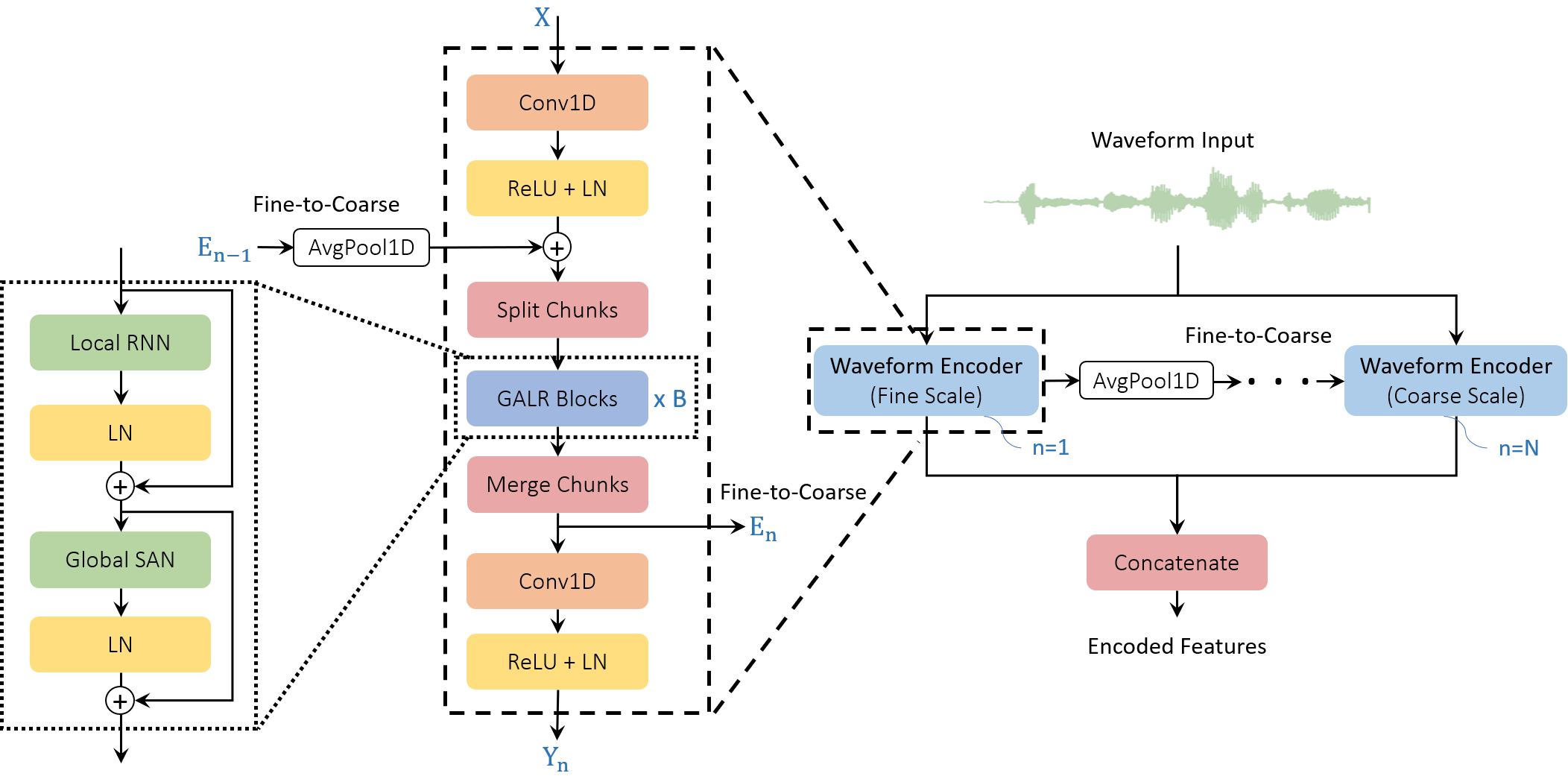}
\vspace{-0.2cm}
\caption{Overall architecture of the multi-scale GALR waveform encoder}
\label{fig:galr}
\vspace{-0.1cm}
\end{figure*}

\section{Introduction}
End-to-end automatic speech recognition (ASR) \cite{graves2012sequence, soltau2017neural, chan2016listen, rao2017exploring, chiu2018state} allows a joint optimization of components for better performances. Amongst a variety of end-to-end methods, RNN-Transducer (RNN-T) \cite{graves2012sequence} has recently achieved state-of-the-art (SOTA) performances \cite{chiu2018state, wang2019exploring, li2019improving, weng2019minimum, sainath2019two, gulati2020conformer}. An RNN-T comprises an encoder-decoder architecture, where the encoder is responsible for transforming a sequence of acoustic inputs into hidden vectors and the decoder utilizes previously emitted non-blank symbols to compute prediction vectors.
A recent study \cite{shrivastava2021echo} suggests that for an RNN-T optimizing encoder and learning proper representations for acoustic inputs are remarkably more vital than the decoder.
\par
Standard ASR systems, including the SOTA RNN-T, traditionally use handcrafted acoustic features computed over segments of fixed duration, such as band-pass filtering of signals including log Mel-filterbank values (FBANK) \cite{Fbank1963} and Mel frequency cepstral coefficients (MFCC) \cite{MFCC1980}.
A potential defect of such systems is being susceptible to information loss due to an intrinsic tradeoff between temporal and frequency resolution. For example, FBANK can hardly preserve phonetic information at the frame boundaries at a fixed time resolution, especially for fast speech, nor describe or separate subtle harmonics and formants at a fixed frequency resolution; Mel-filterbank operations are prone to discard local correlations within the signal \cite{oglic2020deep}.

An increasing number of studies \cite{sainath2013learning, zhu2016learning, zeghidour2018learning, zeghidour2018end, HNey2018, ravanelli2018speech, khan2018learning, loweimi2019learning, oglic2020deep} report empirical evidence that learning directly from the waveform domain can potentially outperform the analytically handcrafted features. In the majority of the previously proposed architectures, there are typically shallow or deep convolutional layers with either 1D convolutions \cite{zhu2016learning, zeghidour2018learning, HNey2018, khan2018learning} or 2D convolutions \cite{oglic2020deep}, designed to emulate band-pass filtering or spectrogram features.
While convolutional networks are good at modeling local correlations as well as invariance to local translations, they can not capture long context with shallow layers. Even with very deep layers \cite{oglic2020deep} at increased computational and memory cost, they are still limited in capturing dynamic long-range global context as CNN simply applies a global averaging over the entire sequence. Existing models using convolutional, recurrent neural networks (RNNs, e.g., LSTMs\cite{gers1999learning} and GRUs \cite{chung2014empirical}), or self-attention networks (SAN) each has its shortcomings: SAN, on the other hand, are well suited to model longer global context but less capable to extract fine-scale local features. 
\par
Grasping the fact that RNN/CNN and SAN are complementary in modeling local detail dependencies and global long-range context, we propose to adopt an architecture called Globally Attentive Locally Recurrent (GALR) network. Its advantage has been demonstrated in the Cocktail-Party problem (i.e., speech separation tasks) in our previous studies \cite{lam2021effective, lam2021sandglasset, wang2021tune}. By a large margin, it outperforms other advanced audio separation networks using a fully convolutional network \cite{luo2019conv} and using a dual-path RNN network \cite{luo2020dual}. Built on the previous success, the unique contributions of this paper are the followings: 
\begin{enumerate}[a)]
\item We reveal the effectiveness of the GALR networks as building modules for a waveform encoder in end-to-end ASR. One challenge in directly taking raw waveform as input is that the sequence is densely sampled in time, and thus presents the dilemma between reducing the number of timesteps and throwing away relevant information for ASR. Our work elegantly solves the critical bottleneck of self-attention networks for modeling very long sequences and makes it memorably and computationally tractable. GALR is well suited to capture not only  long-range global context but also local correlations and fine-grained patterns.
\item We propose a novel architecture that jointly learns GALRs at multiple scales on raw waveforms simultaneously. The combined multi-scale features remedy the tradeoff of temporal and frequency resolution, thus allowing the model to automatically select and combine the resolutions that could most efficiently represent the needed frequencies. Meanwhile, unlike other multi-scale approaches, which generally require very deep structure at extra cost \cite{zhu2016learning}, our method is computationally and memorably efficient.
\item We demonstrate that our waveform encoder can outperform by a large margin over statically hand-engineered features on large vocabulary end-to-end ASR tasks. In particular, we build our baseline models using the RNN-T architecture with SOTA encoders, including Conformer \cite{gulati2020conformer} and TDNN-Conformer \cite{guo2020recent}. We observe a significant gain measured with a benchmark dataset can be consistently carried over to larger setups and generalize well to unseen complex scenarios with highly non-stationary interference. Our evaluations provide evidence for the effectiveness of directly learning a waveform encoder based on GALR and the robustness of withstanding signal corruption, as opposed to building end-to-end ASR on top of traditional features.
\end{enumerate}

\section{Multi-Scale GALR Waveform Encoder}
\label{sec:2}
We propose a raw waveform encoder with a multi-scale globally attentive locally recurrent (GALR) architecture. As shown in Figure \ref{fig:galr}, the waveform input goes through a series of $N$ modules, each composed of $B$ GALR blocks, to encode features at different time scales. 
The resultant features then concatenate and can substitute the conventional FBANK or MFCC features, and instead of being static, they are adaptive and jointly learned during the end-to-end ASR model training.

\subsection{Waveform Encoder}
\label{subsec:Enc}
Given a raw waveform input $\mathbf{x}\in \mathbb{R}^{\Gamma}$, we first split the sequence into $L_n$ half-overlapping windows, denoted by $\mathbf{x}_{1}, ...,\mathbf{x}_{L_n}$, where $\mathbf{x}_{i}\in\mathbb{R}^{M_n}, i=1,..., {L_n}$, and $M_n$ is the window length changing from small to large while the scale index $n$ goes from $1$ to $N$. These windows are $50\%$ overlapped, and the last one is padded with a zero vector to make $L_n=\lceil 2\Gamma/M_n \rceil$. Then, analogous to traditional short-time Fourier transform (STFT), ${L_n}$ frames are generated by non-linearly projecting $\mathbf{x}_{i}$ onto a $D$-dimension feature space, forming a matrix $\mathbf{F}_n \in\mathbb{R}^{D\times L_n}$:
\begin{align}
    \mathbf{F}_n=\text{LN}(\text{ReLU}(\text{Conv1D}([\mathbf{x}_1,...,\mathbf{x}_{L_n}]))),
\end{align}
where $\text{Conv1D}(\cdot)=\mathbf{U}_n*[\mathbf{x}_1,...,\mathbf{x}_{L_n}]$,
$*$ denotes the 1D convolution operation, $\mathbf{U}_n\in \mathbb{R}^{D\times M_n}$ is a learnable matrix of $D$ basis vectors, $\text{ReLU}(\cdot)$ is the rectified linear unit for ensuring the non-negativity \cite{lam2021effective}, and $\text{LN}(\cdot)$ is the layer normalization method \cite{ba2016layer} for stabilizing and accelerating the model training.
\par
To inherit fine-to-coarse information, we then connect the output of the $(n-1)$-th network to $\mathbf{F}_n$:
\begin{align}
\label{eq:avgpool}
\mathbf{\Tilde{F}}_n=\mathbf{F}_n+\text{AvgPool1D}\left(\mathbf{E}_{n-1}\right),
\end{align}
where $\mathbf{E}_{n-1}\in\mathbb{R}^{D\times L_{n-1}}$ is the output (detailed in Section \ref{sec:merge}) of the preceding GALR block at a finer scale (i.e., $L_{n-1} > L_n$), and for initialization, we define $\mathbf{E}_0=\mathbf{0}$ for the first block, and $\text{AvgPool1D}(\cdot)$ is an averaging pooling operation that interpolates the length $L_{n-1}$ of $\mathbf{E}_{n-1}$ to match the length $L_{n}$ of $\mathbf{F}_n$.  

Before conducting GALR operations, we further split $\mathbf{\Tilde{F}}_n$ into $S_n$ half-overlapping chunks each of length $K_n$. The first and last chunks are padded with zeros to generate $S_n=\lceil 2L_n/K_n \rceil$ chunks. These chunks can be represented as a 3D tensor $\mathcal{Q}_n \in \mathbb{R}^{D\times S_n \times K_n}$. 

\subsection{GALR Blocks}
\label{subsec:GALR}
As shown in the middle of Figure \ref{fig:galr}, for the building blocks in each waveform encoder, we adopt the same GALR structure originally proposed for our speech separation tasks \cite{lam2021effective}. In this work, we find that it can generalize surprisingly well for ASR. For integrity, we briefly describe the GALR architecture here. The left diagram in Figure \ref{fig:galr} illustrates the connections within a GALR block. For each block, we denote the input as $\mathcal{Q}_{n,b} \in \mathbb{R}^{D\times S_n \times K_n}$, where $n = 1,..., N$ is the scale index, $b = 1,..., B$ is the block index, and $\mathcal{Q}_{n,1}=\mathcal{Q}_n$. Each GALR block contains two primary layers -- a local RNN layer and a global SAN \cite{vaswani2017attention} layer, corresponding to the intra- and the inter-chunk processing, respectively. 

\subsubsection{Local RNN Layer}
The local RNN layer with $D$ hidden nodes captures short-term dependencies within each chunk:
\begin{align}
    \mathcal{L}_{n,b} = \left[\text{Linear}\left(\text{RNN}\left(\mathcal{Q}_{n,b}[:, s, :]\right)\right), s = 1, . . . , S_n\right],
\end{align}
where $\text{Linear}(\cdot)$ is a trainable linear mapping with a bias term enabled, $\text{RNN}(\cdot)$ is an RNN model, e.g. GRU or LSTM, and $\mathcal{Q}_{n,b}[:, s, :]\in \mathbb{R}^{D\times K_n}$ refers to the local sequence within the $s$-th chunk, each of length $K_n$. The output of the local RNN then goes through $\text{LN}(\cdot)$ followed by a residual connection:
\begin{align}
    \mathcal{\Tilde{L}}_{n,b} = \text{LN}(\mathcal{{L}}_{n,b})+ \mathcal{Q}_{n,b}.
\end{align}

\subsubsection{Global SAN Layer}
Then, we transpose the 3D tensor $\mathcal{\Tilde{L}}_{n,b} \in \mathbb{R}^{D\times S_n \times K_n}$ to process inter-chunk sequences, each of length $S_n$. A multi-head SAN layer \cite{vaswani2017attention} is used to capture the global dependencies:
\begin{align}
\label{eq:3}
    \mathcal{G}_{n,b} = \text{UpSmpl}\left(\text{SAN}\left(\text{DownSmpl}\left(\Tilde{\mathcal{L}}_{n,b}\right)\right)\right),
\end{align}
where $\text{SAN}(\cdot)$, $\text{UpSmpl}(\cdot)$, and $\text{DownSmpl}(\cdot)$ are the self-attentive, upsampling, and downsampling operations, respectively. Readers are referred to \cite{lam2021sandglasset} for more details of these operations, which we omit here for simplicity.
\par
Afterward, the output of the $b$-th GALR block is passed forward with a residual connection as the input for the $(b+1)$-th GALR block as follows: 
\begin{align}
    \mathcal{Q}_{n,b+1} = \text{LN}(\mathcal{{G}}_{n,b})+ \mathcal{\Tilde{L}}_{n,b},
\end{align}

\subsection{Merging Chunks}
\label{sec:merge}
As shown in the middle diagram of Figure \ref{fig:galr}, after processing through the $B$ GALR blocks, we merge the chunks to transform the 3D tensor back to a sequence of feature vectors for the downstream ASR task. Inspired by the mask estimation in \cite{lam2021sandglasset}, the merging operations entail a non-linear mapping, without which we observed a loss degradation in our ablation study. In particular, we use a Swish-gated 2D convolutional layer for the non-linear mapping:
\begin{align}
    \mathcal{F}_n=\text{Conv2D}\left(\text{Swish}\left(\mathcal{Q}_{n,B}\right)\right),
\end{align}
where $\text{Conv2D}(\cdot)$ is a point-wise 2D convolutional layer with a $1\times 1$ kernel, and $\text{Swish}(\cdot)$ is the element-wise Swish activation function \cite{ramachandran2017searching}. Afterward, we merge the $S_n$ chunks using $\text{OverlapAdd}$
\cite{luo2020dual} to transform the 3D tensor $\mathcal{F}_n\in\mathbb{R}^{D\times S_n \times K_n}$ back to a feature matrix:
\begin{align}
    \mathbf{E}_n &=\text{OverlapAdd}\left(\mathcal{F}_n\right),
\end{align}
which is also passed to the next GALR encoder to establish the fine-to-coarse information flow, as indicated in Eq. \ref{eq:avgpool}. 

\subsection{Multi-scale Features}
Current SOTA ASR systems (e.g., RNN-T \cite{gulati2020conformer}) typically use analytical acoustic features computed over a fixed window size of 25ms and a stride length of 10ms. 
Consequently, the time and frequency resolution have been fixed a priori rather than being adaptive or learnable using the training data and the end-to-end model learning. These irreversible operations inevitably cause information loss at a landing stage of scaffold for ASR. 

In the light of the successful fine-grained modeling of GALR networks, we are able to flexibly and effectively handle frames over fine-to-coarse windows (i.e. using 0.2ms to 2ms windows for the Cocktail-Party problem \cite{lam2021effective}). The fine-to-coarse range could be larger in ASR. For example, striking a balance between computation and performance in our practical setup for large industrial datasets, we empirically set three window sizes: 6.25ms, 12.5ms, and 25ms.
During model training we let the network decide which elements under the varying resolutions contain useful information adaptively to the training data based on the back-propagated RNN-T loss.
\par
Optimizing the RNN-T loss per sample \cite{graves2012sequence} entails computing a tensor in shape $D' \times T \times U$, where $T$ is the length of encoded vectors, $U$ is the length of prediction vectors, and $D'$ is the feature dimension. The RNN-T encoder (e.g. in \cite{gulati2020conformer}) has down-sampling processes to keep $T$ small enough for not exceeding GPU memory
and make the back-propagation feasible. Similarly, we add a non-linear down-sampling layer after the chunks-merging operation:
\begin{align}
    \mathbf{Y}_n=\text{LN}\left(\text{ReLU}\left(\text{Conv1D}_{C_n}\left(\mathbf{E}_n\right)\right)\right),
\end{align}
where $\text{Conv1D}_{C_n}(\cdot)$ denotes the 1D-convolution for downsampling with a kernel size of $2C_n$, a stride length of $C_n$, and a padding length of $C_n$ on both sides, such that the resultant length of $\mathbf{Y}_n$ is about $1/C_n$ of the length of $\mathbf{E}_n$, i.e., $\mathbf{Y}_n\in\mathbb{R}^{D\times \lceil L_n/C_n\rceil }$.

After computing the downsampled features $\mathbf{Y}_n$ at $n=1,...,N$ different scales, a straightforward approach to generate a multi-scale feature is to directly concatenate $\mathbf{Y}_n$ along the $D$ dimension. However, there could be a sequence length mismatch problem. Mathematically, we want to make
\begin{equation}
\label{eq:ceil}
\begin{aligned} 
    \left\lceil\frac{L_n}{C_n}\right\rceil=\left\lceil\frac{L_{n+1}}{C_{n+1}}\right\rceil\Leftrightarrow
    \left\lceil\left\lceil\frac{2\Gamma}{M_n}\right\rceil\frac{1}{C_n}\right\rceil&=\left\lceil\left\lceil\frac{2\Gamma}{M_{n+1}}\right\rceil\frac{1}{C_{n+1}}\right\rceil \\
    &{n=1,...,N-1,}
\end{aligned}
\end{equation}
therefore, we set the hyperparameters $M_n$ and $C_n$ to satisfy a relation
$M_{n}C_n=M_{n+1}C_{n+1}.$
\par
Due to the nested ceiling operations in Eq.\ref{eq:ceil}, there could still be cases where some of the $N$ sequences are one element longer or shorter. We define $T=\min_{n} \left\lceil L_n/C_n\right\rceil$ and discard the edge elements indexed at $L>L_\text{min}$. This yields the final multi-scale features:
\begin{align}
\mathbf{Y}=\text{Concat}\left(\mathbf{Y}_1, ..., \mathbf{Y}_n\right).
\end{align}

The encoded waveform features $\mathbf{Y}\in\mathbb{R}^{ND\times T}$ are then fed to the later RNN-T encoder model, e.g., Conformer \cite{gulati2020conformer}.


\section{Evaluation and Analysis}
\label{sec:3}
\subsection{Experimental Setup}

\subsubsection{Data Preparation}
Our models were trained on three datasets at 16kHz sampling rate: the 1,000-hour \textit{AISHELL-2} benchmark \cite{du2018aishell}, a $\sim$5,000-hour (\textit{5khrs}) and a $\sim$21,000-hour (\textit{21khrs}) Mandarin speech datasets \cite{Weng2018}. The models were evaluated on a variety of test sets, covering 1) \textit{Mic}: the benchmark AISHELL-2 test set, 2) \textit{Read}: 1.5-hour read speech (Read), 3) \textit{Spon}: 2-hour spontaneous speech, and 4) \textit{Music}: 2.2-hour speech with background music collected from a real-world video-sharing platform.

\subsubsection{Model Setup and Training Details}
Two cutting-edge encoder architectures that gave SOTA performances \cite{guo2020recent}, Conformer and TDNN-Conformer, were taken as our reference systems. They were built on two types of input representations: 1) Traditional static features. To select the strongest static features, we empirically examined the performances on the largest training set (\textit{21khrs}) with regard to MFCC and FBANK features both following the recipe in Kaldi \cite{povey2011kaldi} (PyKaldi \cite{can2018pykaldi} was used to read the features extracted in Kaldi). On the \textit{Spon} test set, 40-dimension high-resolution MFCC outperformed 80-dimension FBANK by about 0.1\% absolute reduction on average character error rates (CERs). A similar observation was also made in \cite{Weng2018}. Consequently, we selected MFCC for our baseline systems using handcrafted features. 2) End-to-end learnable features, for which we also trained baselines using the single-scale and multi-scale convolution-based approach \cite{zeghidour2018end, zhu2016learning} to directly learn filterbanks from the raw waveform. All systems were trained following the RNN-T, which was implemented in the PIKA\footnote{https://github.com/tencent-ailab/pika} library \cite{Weng2018, Weng2020}, and conducted on 16 NVIDIA Tesla V100 GPU devices. For simplicity, we refer our readers to \cite{Weng2020} for the detailed RNN-T setups.

For the Conformer models, we reproduced the small, medium, and large Conformer encoders, with setups as described in \cite{gulati2020conformer}, and denoted them as Conf-S, Conf-M, and Conf-L. Moreover, the TDNN-Conformer model (denoted as TDNN-Conf) has shown better performances in \cite{guo2020recent} than TDNN-Transformer \cite{weng2019minimum} and Conformer \cite{gulati2020conformer}. Our evaluation also showed a consistent conclusion.
\par
For the multi-scale GALR networks, we set $N=3, B=1, \{M_1, M_2, M_3\}=\{100, 200, 400\}, \{K_1, K_2, K_3\}=\{48, 24, 12\}, \{C_1, C_2, C_3\}=\{8, 4, 2\}$. The GALR feature dimension ($D$) was set to $128$ for \textit{AISHELL-2} and to $256$ for \textit{5khrs} and \textit{21khrs} datasets, respectively.
For a fair comparison, all models had a comparable number of parameters to its counterpart. We removed the MFCC feature extraction step and the first encoder layers, which were the first $8$/$4$/$3$ conformer layers for the S/M/L Conformer models and $3$ TDNN layers for the TDNN-Conformer models, and replaced them with the multi-scale GALR waveform encoder, while the overall sizes matched the respective baseline models.
\subsection{Results and Discussion}
\subsubsection{Performances on Benchmark AISHELL-2}
\begin{table}[th!]
\centering
\vspace{-0.2cm}
\caption{Performances of MFCC and the multi-scale GALR based SOTA encoders trained on the AISHELL-2 training set and evaluated on the AISHELL-2 test set (Mic)}
\label{tab:aishell}
\vspace{-0.1cm}
\begin{tabular}{l|c|c|c|c}
\specialrule{.13em}{0em}{0em} 
\multirow{2}{*}{\bf Encoder} & \multicolumn{2}{c|}{\bf \#Params. (M)} & \multicolumn{2}{c}{\bf CERs (\%)} \\ \cline{2-5}
& \bf MFCC & \bf GALR & \bf MFCC & \bf GALR\\
\hline
Conf-S \cite{gulati2020conformer} & 8.7 & 8.3 & 17.1 & \bf 12.3 \\
Conf-M \cite{gulati2020conformer} & 27.2 & 26.9 & 15.2 & \bf 11.9 \\
Conf-L \cite{gulati2020conformer} & 114 & 114 & 13.3 & \bf 11.7\\
TDNN-Conf \cite{guo2020recent} & 102 & 99.0 & 12.7 & \bf 11.7 \\
\specialrule{.13em}{0em}{0em}
\end{tabular}
\vspace{-0.3cm}
\end{table}
First and foremost, we trained various types of RNN-T encoders on the benchmark \textit{AISHELL-2} \cite{du2018aishell}. As shown in Table \ref{tab:aishell}, across different configurations of the SOTA Conformer encoders, applying the multi-scale GALR waveform encoder in place of the handcrafted features consistently obtained CER reductions, from $7.9\%$ to $28.1\%$ relative. The results also indicated that the traditional TDNN-Conformer outperformed the traditional Conformers, echoing the report by \cite{guo2020recent}, therefore, we selected the TDNN-Conformer encoder as the reference models for the following evaluations in larger scales. 
\par
It was also worth mentioning that, to focus on the comparisons concerning different encoder architectures, all the ASR performances through out this paper were obtained from a pure RNN-T system, without using any additional resource or technique (e.g., NNLM, MBR \cite{weng2019minimum} or LAS rescoring \cite{sainath2019two}), which could be used together with our approach to further boost the performances.
\subsubsection{Performances on 5khrs Using Varying Window Scales}
\begin{table}[t!]
\centering
\caption{Performances of MFCC, Conv1D, and GALR based TDNN-Conformer trained on the \textit{5khrs} speech and evaluated on the Read test set.}
\label{tab:5khrs}
\vspace{-0.1cm}
\begin{tabular}{l|c|c}
\specialrule{.13em}{0em}{0em} 
\bf Features & \makecell{\bf Window Scales} ($M_n$ in ms) & \makecell{\bf CERs (\%)}\\
\hline
MFCC & 25ms & 9.4 \\
\hline
\multirow{3}{*}{Conv1D \cite{zeghidour2018learning, zhu2016learning}} & 25ms & 9.2 \\
 & 12.5ms & 9.1 \\
 & \{6.25,12.5,25\}ms & 8.9\\
\hline
\multirow{4}{*}{GALR} & 12.5ms & 9.0 \\
& \{6.25,12.5\}ms & 7.9 \\
& \{6.25,12.5,25\}ms & 7.6 \\
& \{6.25,12.5,25,50\}ms & \bf 7.4 \\
\specialrule{.13em}{0em}{0em}
\end{tabular}
\vspace{-0.3cm}
\end{table}

This section describes an ablation study to verify that the proposed GALR waveform encoder benefits from the multi-scale structure. To this end, we trained the GALR based TDNN-Conformer in different sets of window scales, as listed with $M_n$ in milliseconds in Table \ref{tab:5khrs}. Here we added another strong baseline that also learned features directly from the raw waveform, i.e. the 1D-convolution-based filterbank learning approach (Conv1D) \cite{zeghidour2018learning}. As shown in Table \ref{tab:5khrs}, the performances of multi-scale GALR based RNN-T systems remarkably reduced the CERs over the non-adaptive handcrafted features from $16.0\%$ to $21.3\%$ relatively. In contrast, although the Conv1D based RNN-T systems also improved over the MFCC baseline, the improvements were marginal compared to GALR.
\subsubsection{Performances on 21khrs Large Industrial Dataset}
\begin{table}[th!]
\centering
\vspace{-0.2cm}
\caption{Performances of the TDNN-Conformer and the one using multi-scale GALR encoder, both trained on the 21khrs speech and evaluated on Read, Spon, and Music test sets}
\label{tab:21khrs}
\vspace{-0.1cm}
\begin{tabular}{l|c|c|c|c}
\specialrule{.13em}{0em}{0em} 
\multirow{2}{*}{\bf Encoder} & \multicolumn{3}{c|}{\bf CERs (\%)} & \multirow{2}{*}{\bf Speed} \\ \cline{2-4}
 & \makecell{\textit{Read}} & \makecell{\textit{Spon}} & \makecell{\textit{Music}} & \\
\hline
TDNN-Conf \cite{guo2020recent} & 4.4 & 13.4 & 29.7 & 239 char/s \\
(w) GALR encoder & \bf 4.1 & \bf 13.2 & \bf 25.2 & \bf 382 char/s \\
\specialrule{.13em}{0em}{0em}
\end{tabular}
\vspace{-0.3cm}
\end{table}

We further investigated the performances on the \textit{21khrs} training set, which was a larger and more realistic industrial dataset. We also used a variety of test sets with a broader diversity such as different speaker speaking styles (i.e., read and spontaneous) and scenarios (with music interference in a video-sharing platform). As shown in Table \ref{tab:21khrs}, the model using the proposed multi-scale GALR encoder obtained consistently lower CERs than the traditional TDNN-Conformer. The result showed that the advantage of our approach could sustain in large-scale data and generalize well to complex scenarios. Especially, the CER for \textit{Music} was reduced by a large margin of $15.2\%$ relatively. Please note that our baseline model was fairly strong using TDNN, which is successful in modeling context information. This demonstrated the powerful capability of the proposed GALR to capture global dynamic context, such as in the unseen complex scenarios with non-stationary music interference as illustrated here. This evaluation empirically proves the robustness of withstanding interference by directly learning a waveform encoder based on GALR. Meanwhile, our approach sped up the processing notably from $239$ to $382$ characters per second.

\section{Conclusions}
\label{sec:4}
We propose a novel architecture that jointly learns GALR at multiple different scales on raw wave-forms simultaneously, which intrinsically remedy the tradeoff of temporal and frequency resolution, thus mitigating the information loss for end-to-end ASR.
Our evaluations demonstrate the effectiveness of directly learning waveform encoder based on GALR, as opposed to building end-to-end ASR on top of traditional features. Moreover, our method is computationally and memory efficient. Future work includes exploring this novel architecture in a cocktail party scenario to solve the speech separation and ASR tasks simultaneously in an end-to-end solution.
\bibliographystyle{IEEEtran}

\bibliography{mybib}


\end{document}